\documentclass[]{interact}

\usepackage{epstopdf}
\usepackage[caption=false]{subfig}

\usepackage[numbers,sort&compress]{natbib}
\bibpunct[, ]{[}{]}{,}{n}{,}{,}
\renewcommand\bibfont{\fontsize{10}{12}\selectfont}
\makeatletter
\def\NAT@def@citea{\def\@citea{\NAT@separator}}
\makeatother

\theoremstyle{plain}

\theoremstyle{definition}

\theoremstyle{remark}

\usepackage[linesnumbered,ruled]{algorithm2e}

\begin{document}
	

\title{Obtaining the mean fields with known Reynolds stresses at steady state}

\author{
	\name{
		Xianwen Guo\textsuperscript{a}
		and Zhenhua Xia\textsuperscript{b}\thanks{CONTACT Zhenhua Xia. Email: xiazh@zju.edu.cn}
		and Heng Xiao\textsuperscript{c}
		and Jinlong Wu\textsuperscript{c}
		and Shiyi Chen\textsuperscript{d,a}}
	\affil{\textsuperscript{a}State Key Laboratory for Turbulence and Complex Systems, College of Engineering, Peking University, Beijing 100871, China;
		   \textsuperscript{b}Department of Engineering Mechanics, Zhejiang University, Hangzhou 310027, China;
		   \textsuperscript{c}Department of Aerospace and Ocean Engineering, Virginia Tech, Blacksburg, VA 24060, USA;
		   \textsuperscript{d}Department of Mechanics and Aerospace Engineering, Southern University of Science and Technology, Shenzhen 518055, China}
}
\maketitle

\begin{abstract}
With the rising of modern data science, data--driven turbulence modeling with the aid of machine learning algorithms is becoming a new promising field. Many approaches are able to achieve better Reynolds stress prediction, with much lower modeling error ($\epsilon_M$), than traditional RANS models but they still suffer from numerical error and stability issues when the mean velocity fields are estimated using RANS equations with the predicted Reynolds stresses, illustrating that the error of solving the RANS equations ($\epsilon_P$) is also very important. In the present work, the error $\epsilon_P$ is studied separately by using the Reynolds stresses obtained from direct numerical simulation and we derive the sources of $\epsilon_P$. For the implementations with known Reynolds stresses solely, we suggest to run an adjoint RANS simulation to make first guess on $\nu_t^*$ and $S_{ij}^0$. With around 10 iterations, the error could be reduced by about one-order of magnitude in flow over periodic hills. The present work not only provides one robust approach to minimize $\epsilon_P$, which may be very useful for the data-driven turbulence models, but also shows the importance of the nonlinear part of the Reynolds stresses in flow problems with flow separations.
\end{abstract}


\section{Introduction}

Turbulence is ubiquitous in nature and engineering applications and it is one of the main research topics in fluid mechanics. Thanks to the rapidly development in computer technology and the numerical algorithm, numerical simulation is becoming a more and more important tool to study turbulence. Although direct numerical simulation (DNS) and large-eddy simulation (LES) can obtain more accurate prediction in turbulence, Reynolds-averaged Navier-Stokes (RANS) is still the most popular simulation approach in engineering design and applications. In RANS simulation, an extra unclosed term, known as the Reynolds stresses, arises due to the nonlinearity of the convective
term in the momentum equation, and thus some treatment, the RANS model, should be adopted to close it~\cite{Chou1940,Kolmogorov1942,Chou1945}.

Let's take the incompressible flow as an example, where the governing equations are as follows:
\begin{eqnarray}
\frac{\partial U_i}{\partial x_i} &=& 0 \label{UEqn1}, \\
\frac{\partial U_i}{\partial t} +U_j\frac{\partial U_i}{\partial x_j} &=& -\frac{1}{\rho}\frac{\partial p}{\partial x_i}+\nu\frac{{\partial}^2 U_i}{\partial x_j\partial x_j}-\frac{\partial R_{ij}}{\partial x_j}. \label{UEqn2}
\end{eqnarray}

\begin{figure}
	\centering
	\includegraphics[scale=0.65]{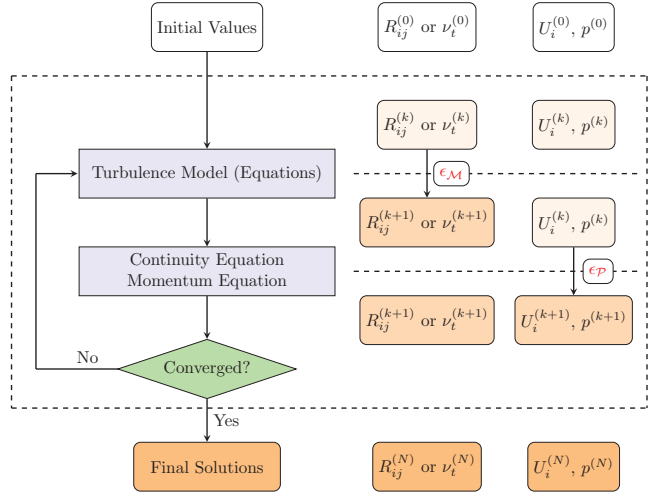}
	\caption{An overview of processes for a typical RANS solver.}
	\label{fig:SegregatedSolvers}
\end{figure}

Here $R_{ij} = \left\langle u_i u_j\right\rangle$ is the unclosed Reynolds stress tensor.
Thanks to the continuous effort by the turbulence community, many different types of models have been proposed for RANS simulations, either with the Boussinesq assumption (Algebraic models or zero-equation models~\cite{BL1978}, one-equation models~\cite{SA} and two equation models~\cite{kEpsilon,LaunderSharma,kOmega}) or beyond it (Stress-transport models~\cite{LRR,SSG} and nonlinear models~\cite{Pope1975})~\cite{Speziale1991,Wilcox2006,Durbin2018}. As sketched in Figure~\ref{fig:SegregatedSolvers}, two different sources of errors could exist for a typical RANS simulation. One is the model error $\epsilon_M$, which comes out when $R_{ij}$ is estimated through the RANS models and it can be denoted as $\epsilon_M=f(R^T_{ij}-R^M_{ij})$ with $R^T_{ij}$ and  $R^M_{ij}$ being the true Reynolds stresses and the modelled Reynolds stresses respectively. The other is the numerical error during the propagation process $\epsilon_P$, which appears when the RANS governing equations \eqref{UEqn1} and \eqref{UEqn2} are solved with RANS closure models inserted. In the past, when a RANS model is evaluated in \textit{a posterior} tests, the final mean fields $U_i^N$ will be compared to the reference true values $U_i^T$ and the deviations can be separated into two parts, i.e.
\begin{equation}\label{Devia}
U_i^T-U_i^N = (U_i^T-U_i^M) + (U_i^M-U_i^N).
\end{equation}
where the first part $(U_i^T-U_i^M)$ is caused by model error $\epsilon_M$ and the second part $(U_i^M-U_i^N)$ is caused by the error $\epsilon_P$. Due to the coupling of the two errors, very little attention was paid to $\epsilon_P$ alone in the previous studies.

With the Boussinesq assumption, the stability is generally not a big issue when the RANS governing equations are solved. However, it becomes much severer if a RANS model beyond the Boussinesq assumption is considered, and convergent solutions may not been obtained at some situations~\cite{Wu2019b, zhao2019turbulence}, making $\epsilon_P$ an important issue that needs to be treated seriously.

Recently, data-driven turbulence modeling has been becoming a promising research field, and many different RANS models have been proposed with the help of different machine learning algorithm~\cite{Parish2016,Sandberg2016,Ling2016b,Duraisamy2017,Wang2017,Wu2018,Zhu2019,AnnualReview,Fang2019,Pandey2020}.
For most data-driven RANS models, no explicit expressions for the Reynolds stress tensor can be obtained~\cite{Ling2016b,Wang2017,Wu2018}, and the numerical instability is even severer and $\epsilon_P$ could be very large. In Ref.~\cite{Wang2017}, they reported that the mean velocity field obtained with their data-driven RANS model does not match better with the DNS data than the original RANS model, even though their RANS model can predict better Reynolds stresses. In Ref.~\cite{Wu2019b}, they believed that RANS equations with explicit data-driven RANS models can be ill-conditioned. In order to make the RANS simulations more stable, they proposed an implicit treatment. With the information of the strain-rate tensor from the DNS database, this implicit treatment can also reduce $\epsilon_P$ to a very low level. Nevertheless, the consistent and accurate strain-rate tensor is not always known in advance, which limits the usage of this implicit treatment.


On the other hand, it has been shown by Thompson et al.~\cite{Thompson2016} that the error of solving RANS equations with Reynolds stresses from accurate DNS can still be very large. With friction Reynolds number $Re_\tau=5200$ in turbulent channel flow, a $0.41 \%$ maximum error in turbulent shear stresses could finally lead to a $21.6 \%$ volume-averaged error in the mean velocities~\cite{Thompson2016,Wu2019b}. In Ref.~\cite{Ling2016b}, they also reported that their predicted streamwise velocity using true DNS anisotropy behaves differently from that from the true DNS (Figure 5 in Ref.~\cite{Ling2016b}). From these results, we may conclude that $\epsilon_P$ could be very large if the RANS governing equations \eqref{UEqn1} and \eqref{UEqn2} are not solved properly.

The present paper aims to study the propagation error $\epsilon_P$ when the mean flow fields are solved with known Reynolds stresses. The Reynolds stresses obtained from DNS are adopted to minimize the influence of $\epsilon_M$.



\section{Methodology}

\subsection{Implicit treatment with known $R_{ij}^{DNS}$ and $S_{ij}^{DNS}$}

Firstly, let's consider the momentum equation appeared in~\eqref{UEqn2} at steady state.
With the deviatoric anisotropic part of Reynolds stress tensor $a_{ij}=\langle u_iu_j\rangle-2k\delta_{ij}/3$ and an alternative pressure $\tilde{P}=p/\rho+2k/3$, the momentum equation can be rewritten as
\begin{equation}
U_j\frac{\partial U_i}{\partial x_j} = -\frac{\partial \tilde{P}}{\partial x_i}+\nu\frac{{\partial}^2 U_i}{\partial x_j\partial x_j}-\frac{\partial a_{ij}}{\partial x_j}. \label{NS*}
\end{equation}

As shown in Ref.~\cite{Wu2019b}, if the above equation~\eqref{NS*} was directly solved with iterative CFD solvers, the local conditioning number could be very large for the corresponding linear algebraic system, making it very difficult to obtain a stable converged solution for equation~\eqref{NS*}. With known $R_{ij}^{DNS}$ and $S_{ij}^{DNS}$ from the DNS data, Wu et al.~\cite{Wu2019b} proposed an implicit treatment. The basic idea is to decompose $a_{ij}^{DNS}$ into a linear part and a nonlinear part based on eddy-viscosity hypothesis which is written as
\begin{equation}
a_{ij}^{DNS} = -2\nu_t S_{ij}^{DNS}+R_{ij}^{\bot}, \label{tau}
\end{equation}
for incompressible flows. Here,
\begin{equation}
R_{ij}^{\bot} = a_{ij}^{DNS} + 2\nu_t S_{ij}^{DNS} \label{tau_Dec}
\end{equation}
is the nonlinear part of the Reynolds stresses, $S_{ij}^{DNS}=(\partial U_i^{DNS}/\partial x_j + \partial U_j^{DNS}/\partial x_i)/2$ is the mean strain rate tensor from the DNS field,
$\nu_t$ is the effective turbulence eddy-viscosity, which is the key to quantify and balance the amount of Reynolds stress to be treated implicitly. With the above decomposition \eqref{tau}, the equation \eqref{NS*} can be transformed into
\begin{align}
U_j\frac{\partial U_i}{\partial x_j} &=-\frac{\partial \tilde{P}}{\partial x_i}+\frac{\partial}{\partial x_j}\left[\left(\nu+\nu_t\right)\frac{\partial U_i}{\partial x_j}\right]-\frac{\partial R_{ij}^{\bot}}{\partial x_j}. \label{NS_imp}
\end{align}

Interestingly, although equation \eqref{NS_imp} is exactly equivalent to equation \eqref{NS*}, better stability property can be achieved, which can be explained by the smaller local condition numbers as elucidated by Wu \textit{et al.}~\cite{Wu2019b}, when it is solved numerically with some algorithms (such as SIMPLE algorithm) to obtain its solution $U^{I}_i$ as
\begin{equation}
\left(U^{I}_j\frac{\partial}{\partial x_j}-\frac{\partial}{\partial x_j}\left[(\nu+\nu_t)\frac{{\partial}}{\partial x_j}\right]\right)U^{I}_i=-\frac{\partial\tilde{P}}{\partial x_i} - \frac{\partial R^{\bot}_{ij}}{\partial x_j}+E_N. \label{LNI}
\end{equation}
Here, $E_N$ is the numerical error when equation \eqref{NS_imp} is solved, which depends on the numerical schemes, the grid used, the algorithm used to solve the algebraic system and so on.
Equivalently, the above equation~\eqref{LNI} can be reformed as
\begin{equation}
\left(U^{I}_j\frac{\partial}{\partial x_j}-\nu\frac{\partial^2}{\partial x_j\partial x_j}\right)U^{I}_i=-\frac{\partial\tilde{P}}{\partial x_i} - \frac{\partial a_{ij}}{\partial x_j}+E_p \label{LNI-2}
\end{equation}
with
$$E_p=\frac{\partial}{\partial x_j}\left[2\nu_t(S^{I}_{ij}-S^{DNS}_{ij})\right]+E_N$$
is the main source of $\epsilon_P$. Ideally, if converged solution $U^{I}_i$ is obtained and it approaches to $U^{DNS}_i$, $\epsilon_P$ could be eliminated. However, the inconsistence between $U^{DNS}_i$ and $a_{ij}^{DNS}$ as well as the existence of $E_N$ makes $\epsilon_P$ inevitable. A proper choice of $\nu_t$ can reduce $\epsilon_P$ to a relatively low level.

\subsection{Propagation with known $R_{ij}^{DNS}$ and unknown $S_{ij}^{DNS}$}

In the applications of RANS simulations, $R_{ij}$ could generally be obtained through some RANS models while $S_{ij}$ can only be estimated from the current field. With the most accurately estimated $R_{ij}=R_{ij}^{DNS}$ and unknown $S_{ij}^{DNS}$, we still need to find some way to obtain the mean field stably while make $\epsilon_P$ as small as possible.

Similar to the decomposition in \eqref{tau}, $a_{ij}^{DNS}$ can still be decomposed with any other known $S_{ij}^*$ and $\nu_t^*$, as
\begin{gather}
a_{ij}^{DNS} = -2\nu_t^* S_{ij}^{*}+R_{ij}^{\bot*}. \label{tau*}
\end{gather}
and the corresponding $R_{ij}^{\bot*}$ can be further determined through
\begin{gather}
R_{ij}^{\bot*} = a_{ij}^{DNS} + 2\nu_{t}^{*} S_{ij}^{*}. \label{tauN*}
\end{gather}
Since $R_{ij}^{\bot*}$ as well as $S_{ij}^{*}$ can only be determined using the velocity field at the current step, iterations should be adopted to solve the problem, and the numerical solution $U^{n}_i$ at the next time step satisfies
\begin{equation}
\left(U^{n}_j\frac{\partial}{\partial x_j}-\frac{\partial}{\partial x_j}\left[(\nu+\nu_t^*)\frac{{\partial}}{\partial x_j}\right]\right)U^{n}_i=-\frac{\partial\tilde{P}}{\partial x_i} - \frac{\partial R^{\bot*}_{ij}}{\partial x_j} + E^n_N, \label{LNI_3}
\end{equation}
with $R_{ij}^{\bot*} = a_{ij}^{DNS} + 2\nu_{t}^{*} S_{ij}^{n-1}$. Here $E^n_N$ is the numerical error at the $n-th$ step due to the numerical algorithm. Rewriting~\eqref{LNI_3}, we have
\begin{equation}
\left(U^{n}_j\frac{\partial}{\partial x_j}-\nu\frac{\partial^2}{\partial x_j\partial x_j}\right)U^{n}_i=-\frac{\partial\tilde{P}}{\partial x_i} - \frac{\partial a_{ij}}{\partial x_j}+E_p^n \label{LNI-4}
\end{equation}
with
\begin{equation}E_p^n=\frac{\partial}{\partial x_j}\left[2\nu_t^*(S^{n}_{ij}-S^{n-1}_{ij})\right] + E^n_N \label{Eq:Error_EPN}
\end{equation}
being the source of $\epsilon_P$ at the $n-th$ step. The final $\epsilon_P$ will be determined by all $E_p^n$ in the past $n$ steps, accumulatively, making both $\nu_t^*$ and $S_{ij}^0$ very important. Again, a choice of $\nu_t^*$ with larger values can make equation~\eqref{LNI_3} more stable, but it may also increase $E_p^n$.

Since we only have the information of $a_{ij}$ at the current situation, we need to run some adjoint RANS simulation make a first guess on $\nu_t^*$ and $S_{ij}^0$. With the information of $\nu_t^{R}$ and $S_{ij}^{R}$ from the adjoint RANS simulation, we could make some suggestions on $S_{ij}^0=S_{ij}^{R}$ and $\nu_t^*$, either
\begin{equation}
\nu_t^*=\nu_t^{R}, \label{eq:A1}
\end{equation}
or
\begin{equation}
\nu_t^*=-\frac{1}{2} \frac{a_{ij} S_{ij}^{R}}{S_{ij}^{R} S_{ij}^{R}}. \label{eq:A2}
\end{equation}
In the following, the above two choices will be denoted as algorithm A1 and A2 respectively. Details of these two methods are summarized in Algorithm \ref{Imp1} and Algorithm \ref{Imp2}.

\begin{algorithm}[h]
	\caption{Baseline RANS Correction (A1)}
	\label{Imp1}
	Run baseline RANS simulation to obtain $S_{ij}^{RANS}$ and $\nu_t^{RANS}$ \;
	Set $\nu_t^*=\nu_t^{RANS}$, $S_{ij}^*=S_{ij}^{RANS}$ \;
	Obtain $R_{ij}^{\bot*}$ through \eqref{tauN*} \;
	\For{$(k=0; k<N; k++)$}
	{
		Solve equation \eqref{LNI_3} to get intermediate velocity fields $U_i^{(k)}$ \;
		Calculate $S_{ij}^{(k)}$ through $U_i^{(k)}$ \;
		Update $R_{ij}^{\bot*} = a_{ij}^{DNS} + 2\nu_{t}^{RANS} S_{ij}^{(k)}$
	}
\end{algorithm}

\begin{algorithm}[h]
	\caption{Maximum Linearization (A2)}
	\label{Imp2}
	Run baseline RANS simulation to obtain $S_{ij}^{RANS}$ and $\nu_t^{RANS}$ \;
	Set $S_{ij}^*=S_{ij}^{RANS}$ \;
	Obtain $\nu_t^*$ through \eqref{eq:A2} \;
	Obtain $R_{ij}^{\bot*}$ through \eqref{tauN*} \;
	\For{$(k=0; k<N; k++)$}
	{
		Solve equation \eqref{LNI_3} to get intermediate velocity fields $U_i^{(k)}$ \;
		Calculate $S_{ij}^{(k)}$ through $U_i^{(k)}$ \;
		Update $\nu_{t}^{m*} = -\frac{1}{2} \frac{a_{ij}^{DNS} S_{ij}^{(k)}}{S_{ij}^{(k)} S_{ij}^{(k)}}$ \;
		Update $R_{ij}^{\bot*} = a_{ij}^{DNS} + 2\nu_{t}^{m*} S_{ij}^{(k)}$ \;
	}
\end{algorithm}

\section{Numerical results}

\begin{figure}
	\centering
	\includegraphics[width=0.6\textwidth]{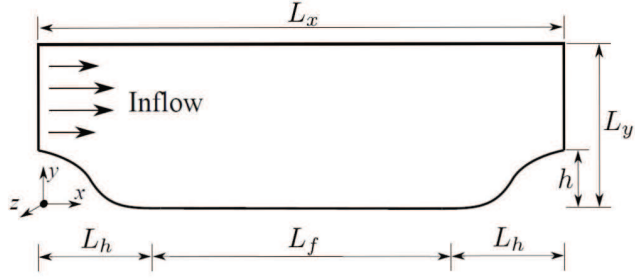}
	\caption{A sketch of computational domain for flow over 2D periodic hills. The hill height is denoted by $h$. The total length and height of the whole domain are denoted by $L_x$ and $L_y$ respectively.}
	\label{fig:sketch}
\end{figure}

In this section, the above proposed two different algorithms will be tested numerically in the flow over two-dimensional (2D) periodic hills, where flow separation and reattachment occur on a smooth curved boundary surface. Due to its relatively simple geometry and well-defined boundary conditions~\cite{Almeida1993, Mellen2000, Breuer2009, Xiao2019}, it has often been used as benchmark test cases for modeling and simulation issues, such as subgrid-scale models and wall functions in LES~\cite{Mellen2000, Temmerman2003, Xia2013}, data-driven turbulence modeling~\cite{Wang2017,Wu2018,Wu2019a}. A sketch of basic geometry is shown in Figure ~\ref{fig:sketch}. The streamwise and vertical directions are denoted as $x$ and $y$ respectively.  The Reynolds number is defined based on the hill height $h$ and the bulk velocity $U$ at inflow section, $Re=Uh/\nu$ with $\nu$ the kinematic viscosity. The hill length is denoted by $L_h$ and the length of flat part of bottom wall is denoted by $L_f$. An accurate specification for hill shape is available in form of piece-wise polynomials in~\cite{Almeida1993, Xiao2019}.

A 2D structured grid is adopted with resolution $128\times160$ in streamwise and normal direction. The grid is refined in the near wall region to ensure that the height of the first cell center above the wall in wall unit is less than 1 for all cases and the grid independence has been checked. All RANS simulations are conducted via the steady-state solver "simpleFoam" based on SIMPLE algorithm (for Semi-Implicit Method for Pressure-Linked Equations)~\cite{Caretto1973} from the widely used open-source platform OpenFOAM \cite{OpenFOAM}. The flow is set to be periodic in the streamwise direction. No-slip condition and zero-gradient condition are set at walls for velocity field and pressure respectively.

As the first test, we would like to show the results of the two algorithms at $Re=10595$ with the Spalart-Allmaras model~\cite{SA} as the adjoint RANS model. The DNS data of Reynolds stress fields are referred to Breuer et al.~\cite{Breuer2009}. The ratio $\sigma\equiv\delta U_{rms}/U_{rms}^{DNS}$ is used to quantify the error of solved mean velocity field to the reference DNS field, where $\delta U_{rms}$ and $U_{rms}^{DNS}$ are defined as~\cite{Wu2019b}
\begin{gather}
\delta U_{rms}=\sqrt{\frac{\sum_{j=1}^{N}\left([U]_{j}-\left[U^{D N S}\right]_{j}\right)^{2}\left[\Delta V_{j}\right]}{V}}\\
U_{rms}^{DNS}=\sqrt{\frac{\sum_{j=1}^{N}\left(\left[U^{DNS}\right]_{j}\right)^{2}\left[\Delta V_{j}\right]}{V}}. \label{UDrms}
\end{gather}
Here, $[\phi]_j$ denotes the $j$-th component of the $N$-vector obtained by discretizing the field $\phi$ on the $N_x\times N_y$ mesh with $N=N_x\times N_y$.

\begin{figure*}
\centering
\includegraphics[width=0.8\textwidth]{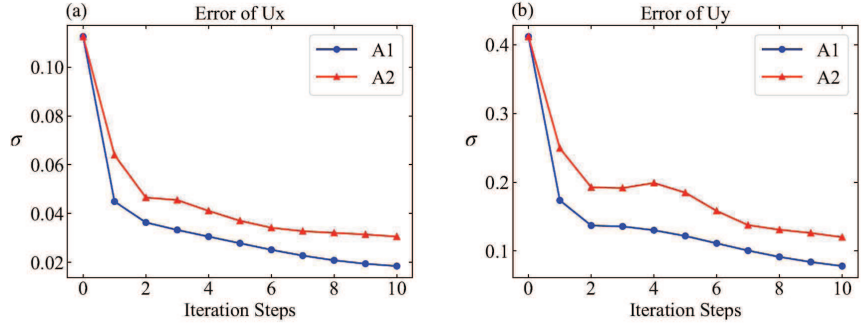}
\caption{\label{fig:ErrPH10595}Errors of propagated velocity for $Re=10595$ at different iteration steps with SA model as the adjoint RANS model via algorithms $A1$ and $A2$: (a) streamwise velocity $U_x$ and (b) vertical velocity $U_y$. The corresponding errors for $U_x$ and $U_y$ using known $S_{ij}^{DNS}$ are about 0.0083 and 0.042 respectively.}
\end{figure*}

\begin{figure*}
	\centering
	\includegraphics[scale=0.4]{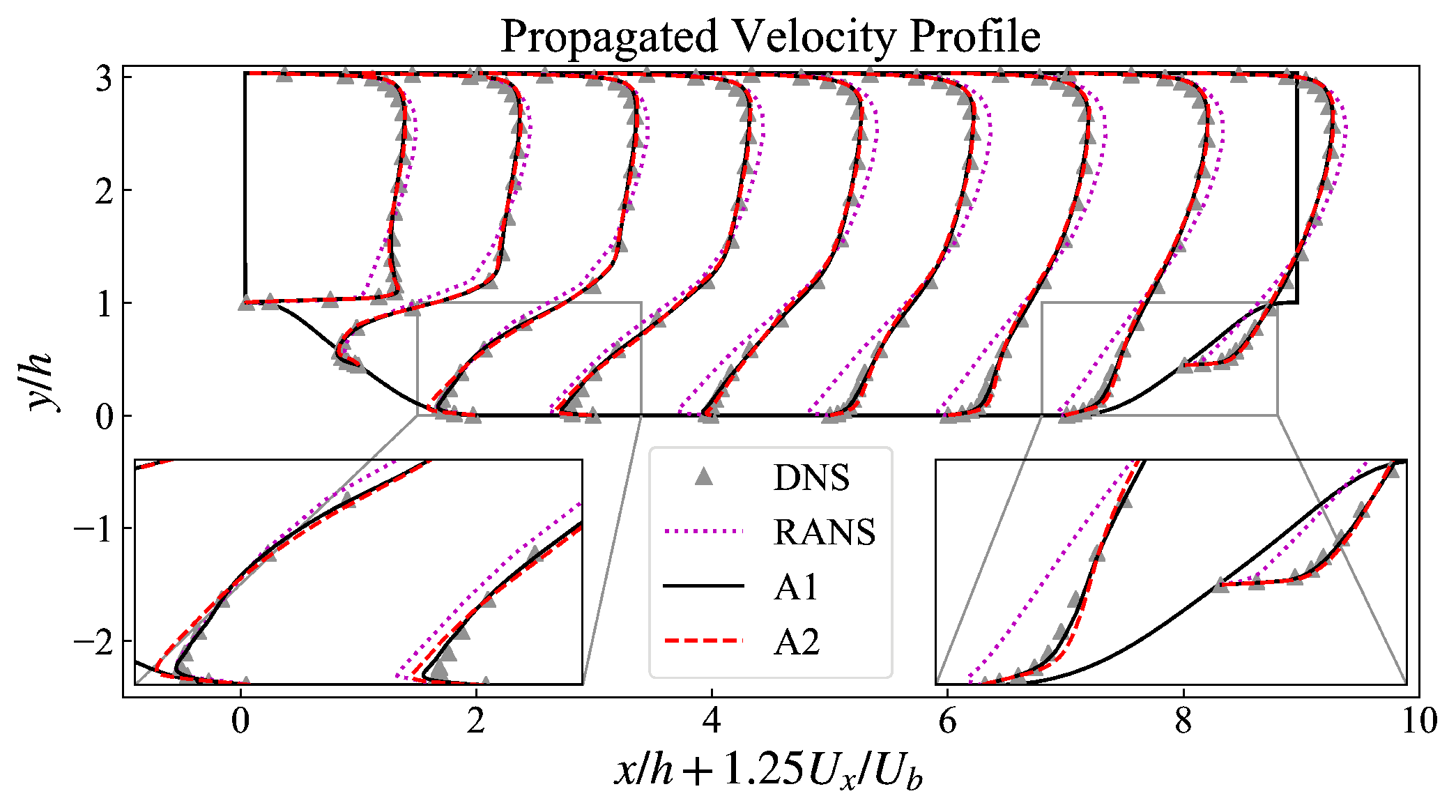}
	\caption{Propagated streamwise velocity profiles on 9 separated locations for $Re=10595$ from the algorithms $A1$ and $A2$. The DNS data and RANS-SA results are also shown for comparison.}
	\label{fig:VelPH10595SA}
\end{figure*}

Figure~\ref{fig:ErrPH10595} shows the errors of streamwise and vertical mean velocities for $Re=10595$ with SA model as the adjoint RANS model via two algorithms as the iteration advances. It is seen that the errors of two velocity components from algorithm $A1$ and $A2$ generally decrease with the iteration advances. They decay very fast during the first two iterations and then become slowly if the iteration goes on. The errors from algorithm $A1$ is lower than those from algorithm $A2$. For $U_x$ and $U_y$, $\sigma$ are about $0.018$ and $0.078$ for algorithm $A1$ while they are about $0.031$ and $0.120$ for algorithm $A2$ after 10 iterations. Comparing to the corresponding errors from the propagation with $S_{ij}^{DNS}$, which are $0.0083$ and $0.042$, the errors for algorithm $A1$ are about twice while those for algorithm $A2$ are about four or three times. Considering that algorithm $A1$ and $A2$ do not need the information about $S_{ij}^{DNS}$, the present increase in errors is acceptable. Furthermore, the errors from baseline RANS, which are about $0.113$ and $0.412$, are one order of magnitude larger than those with correct Reynolds stresses. In order to further show the differences among different results, the mean streamwise velocity profiles from two different algorithms, as well as those from DNS and baseline RANS simulations, at 9 different locations are shown in Figure~\ref{fig:VelPH10595SA}. Firstly, it is seen that the baseline RANS with SA model fails to predict the mean velocity profiles at the 9 locations as compared to the reference DNS profiles. Fictitious backflows can still be observed even at $x/h=6$ for RANS-SA simulation. It is interesting to note that although the baseline RANS predict the mean velocity very poorly, it still can help to promote the prediction of the algorithm $A1$ and $A2$ as shown in Figure~\ref{fig:VelPH10595SA}, where the mean streamwise velocity profiles at 9 different locations match very well with the DNS data. Compared to the baseline RANS simulations, the algorithm $A1$ has the same eddy viscosity besides the additional nonlinear Reynolds stresses. The better prediction on the mean velocities of the algorithm $A1$ over the baseline RANS then well documents the importance of the nonlinear part of the Reynolds stresses in this kind of flow problems with separations and reattachments. The algorithm $A2$ also can get a very good prediction on the mean velocity profiles, although a little poorer than the algorithm $A1$. This results then illustrates that the choice of $\nu_t$ will influence the final results to a certain degree.

\section{Discussions}
\subsection{Influence of adjoint RANS models}

\begin{figure*}
	\centering	\includegraphics[width=0.8\textwidth]{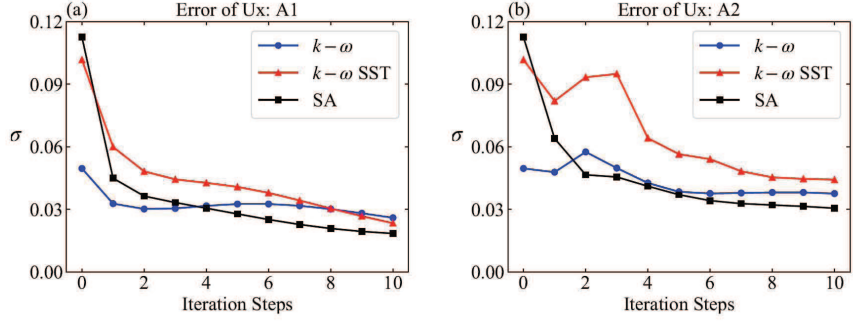}
	\caption{Errors of propagated velocity $U_x$ using $A1$ (a) and $A2$ (b) with different adjoint RANS models at $Re=10595$.}
	\label{fig:ErrTurbs}
\end{figure*}

Since the choice of $\nu_t$ will affect the final results, the different choice of adjoint RANS models will surely affect the final results. Figure~\ref{fig:ErrTurbs} shows the errors of $U_x$
using $A1$ and $A2$ with three different adjoint RANS models, including the SA model,
the $k$-$\omega$ model~\cite{Wilcox2006} and the $k$-$\omega$ SST model~\cite{kOmegaSST}. It is seen that both algorithms $A1$ and $A2$ can effectively reduce the errors to a relatively lower level, as compared to the errors of baseline RANS simulations, i.e. the errors at the $0-th$ iteration, although they are different for different models. Clearly, $A1$ is more stable than $A2$ and its errors are also smaller, again confirms that $A1$ is better than $A2$. We also tested the two algorithms at different geometries and different Reynolds numbers (not shown here), and the results showed that $A1$ is more stable than $A2$, with smaller errors.


\subsection{Other possible approaches}

In the above discussions, we have shown that setting $\nu_t^*=\nu_t^R$ can generally reduce $\epsilon_P$ to a lower level. However, from equations \eqref{LNI-4} and \eqref{Eq:Error_EPN}, other choices could still be adopted. For example, we could set $$\nu_t^*=C\nu_t^R$$ where $C$ is a constant in the whole domain, which can be constant or changing during iterations. For SA model, $C\approx0.7$ could get a slightly better prediction on the mean velocity field. However, we could not determine $C$ in advance without testing which surely will restrict its implementations.

Another choice is to include the history effect when estimates $R_{ij}^{\bot*}$ at the $n-th$ step, such as
$$R_{ij}^{\bot(n)} = a_{ij}^{DNS} + 2\nu_{t}^{*} [S_{ij}^{n} + \alpha (S_{ij}^{n}-S_{ij}^{n-1})].$$
or we could use the adaptive gradient algorithm~\cite{AdaGrad} or ``Adam'' algorithm~\cite{Adam} to make a better estimation on $R_{ij}^{\bot*}$. However, based on our numerical tests, we can only lower the error a little bit.

\section{Conclusions}

In traditional RANS simulations, the propagation error $\epsilon_P$ of obtaining the mean fields with known Reynolds stresses was usually misinterpreted as a part of the modelling error, and has seldom been discussed alone. This makes the judgement on the turbulence models very ambiguous, especially for the data-driven turbulence models which are very popular nowadays. In the present paper, we studied the propagation error $\epsilon_P$ solely by using the Reynolds stresses from DNS databases, and the sources of $\epsilon_P$ was derived for the situations with or without known $S_{ij}^{DNS}$. For general implementations without known $S_{ij}^{DNS}$, the choice of $\nu_t^*$ is very critical. If it is too small, the numerical algorithm may be unstable which will increase the error due to the numerical algorithm. On the other hand, if it is too large, it will increase the iteration errors during two adjacent iterations instead. An adjoint RANS simulation was suggested to make a first guess on $\nu_t^*$ and a good, stable choice is setting $\nu_t^*$ to the eddy viscosity from RANS simulations. With around ten iterations, the error of mean velocity could be reduced by one-order of magnitude. The present work may offer some valuable references for turbulence models beyond the Boussinesq assumption to obtain satisfactory mean velocity fields.

Another outcome of the present work is on the modelling issues. The Algorithm $A1$ can be viewed as a nonlinear correction to the adjoint linear eddy-viscosity RANS model. The better prediction using $A1$ on the mean velocity fields over the baseline RANS model confirms the importance of non-linear part Reynolds stresses, especially for the current type of flow problems with flow separations. This may be helpful for the those groups who are trying to develop advanced data-driven turbulence models.

\section*{Acknowledgement}


Guo, Xia and Chen would like to thank the support by the National Science Foundation of
China (NSFC grant nos. 11822208, 11772297, 91852205 and 91752202). Xia would also like to thank the support from the Fundamental Research Funds for the central Universities.

\section*{Disclosure statement}
No potential conflict of interest was reported by the author(s).

\section*{Funding}

Guo, Xia and Chen is supported by the National Science Foundation of
China (NSFC grant nos. 11822208, 11772297, 91852205 and 91752202). Xia is also supported from the Fundamental Research Funds for the central Universities.



\bibliographystyle{tfnlm}
\bibliography{BackP-RANS}

\begin{thebibliography}{10}
\providecommand{\url}[1]{\normalfont{#1}}
\providecommand{\urlprefix}{Available from: }

\bibitem{Chou1940}
Chou~PY. On an extension of {Reynolds'} method of finding apparent stress and
  the nature of turbulence. Chin J Phys. 1940;\hspace{0pt}4:1--33.

\bibitem{Kolmogorov1942}
Kolmogorov~AN. The equations of turbulent motion in an incompressible fluid.
  Izvestia Acad Sci, USSR; Phys. 1942;\hspace{0pt}6:56--58.

\bibitem{Chou1945}
Chou~PY. On velocity correlations and the solutions of the equations of
  turbulent fluctuation. Quart Appl Math. 1945;\hspace{0pt}3:38--54.

\bibitem{BL1978}
Baldwin~B, Lomax~H. Thin-layer approximation and algebraic model for seperated
  turbulent flows. AIAA Paper 78-257; 1978.

\bibitem{SA}
Spalart~PR, Allmaras~SR. A one-equation turbulence model for aerodynamic flows.
  Rech Aerosp. 1994;\hspace{0pt}1:5--21.

\bibitem{kEpsilon}
Jones~W, Launder~B. The prediction of laminarization with a two-equation model
  of turbulence. Int J Heat Mass Trans. 1972;\hspace{0pt}15(2):301--314.

\bibitem{LaunderSharma}
Launder~B, Sharma~B. Application of the energy-dissipation model of turbulence
  to the calculation of flow near a spinning disc. Lett Heat Mass Trans.
  1974;\hspace{0pt}1(2):131--137.

\bibitem{kOmega}
Wilcox~DC. Reassessment of the scale-determining equation for advanced
  turbulence models. AIAA J. 1988;\hspace{0pt}26(11):1299--1310.

\bibitem{LRR}
Launder~BE, Reece~GJ, Rodi~W. Progress in the development of a reynolds-stress
  turbulence closure. J Fluid Mech. 1975;\hspace{0pt}68(3):537--566.

\bibitem{SSG}
Speziale~CG, Sarkar~S, Gatski~TB. Modelling the pressure-strain correlation of
  turbulence: an invariant dynamical systems approach. J Fluid Mech.
  1991;\hspace{0pt}227:245--272.

\bibitem{Pope1975}
Pope~SB. A more generative effective-viscosity model. J Fluid Mech.
  1975;\hspace{0pt}72:331--340.

\bibitem{Speziale1991}
Speziale~CG. Analytical methods for the development of reynolds-stress closures
  in turbulence. Annu Rev Fluid Mech. 1991;\hspace{0pt}23(1):107--157.

\bibitem{Wilcox2006}
Wilcox~D. Turbulence modeling for cfd. Third edition ed. DCW Industries; 2006.

\bibitem{Durbin2018}
Durbin~P. Some recent developments in turbulence closure modeling.
  Annu~Rev~Fluid~Mech. 2018;\hspace{0pt}50:77--103.

\bibitem{Wu2019b}
Wu~J, Xiao~H, Sun~R, et~al. Reynolds-averaged navier-stokes equations with
  explicit data-driven reynolds stress closure can be ill-conditioned. J Fluid
  Mech. 2019;\hspace{0pt}869:553--586.

\bibitem{zhao2019turbulence}
Zhao~Y, Akolekar~HD, Weatheritt~J, et~al. Turbulence model development using
  cfd-driven machine learning ; 2019.

\bibitem{Parish2016}
Parish~EJ, Duraisamy~K. A paradigm for data-driven predictive modeling using
  field inversion and machine learning. J Comput Phys.
  2016;\hspace{0pt}305:758--774.

\bibitem{Sandberg2016}
Weatheritt~J, Sandberg~R. A novel evolutionary algorithm applied to algebraic
  modifications of the {RANS} stress¨cstrain relationship. J Comput Phys.
  2016;\hspace{0pt}325:22--37.

\bibitem{Ling2016b}
Ling~J, Kurzawski~A, Templeton~J. Reynolds averaged turbulence modelling using
  deep neural networks with embedded invariance. J Fluid Mech.
  2016;\hspace{0pt}807:155--166.

\bibitem{Duraisamy2017}
Duraisamy~K, Singh~AP, Pan~S. Augmentation of turbulence models using field
  inversion and machine learning. In: AIAA SciTech Forum; 55th AIAA Aerospace
  Sciences Meeting; 2017. p. 1--18. 2017-0993.

\bibitem{Wang2017}
Wang~JX, Wu~JL, Xiao~H. Physics-informed machine learning approach for
  reconstructing reynolds stress modeling discrepancies based on dns data. Phys
  Rev Fluids. 2017;\hspace{0pt}2:034603.

\bibitem{Wu2018}
Wu~JL, Xiao~H, Paterson~E. Physics-informed machine learning approach for
  augmenting turbulence models: A comprehensive framework. Phys Rev Fluids.
  2018;\hspace{0pt}3:074602.

\bibitem{Zhu2019}
Zhu~L, Zhang~W, Kou~J, et~al. Machine learning methods for turbulence modeling
  in subsonic flows around airfoils. Phys Fluids.
  2019;\hspace{0pt}31(1):015105.

\bibitem{AnnualReview}
Duraisamy~K, Iaccarino~G, Xiao~H. Turbulence modeling in the age of data. Annu
  Rev Fluid Mech. 2019;\hspace{0pt}51(1):357--377.

\bibitem{Fang2019}
Fang~R, Sondak~D, Protopapas~P, et~al. Neural network models for the
  anisotropic reynolds stress tensor in turbulent channel flow. Journal of
  Turbulence. 2019;\hspace{0pt}0(0):1--19.
  \urlprefix\url{https://doi.org/10.1080/14685248.2019.1706742}.

\bibitem{Pandey2020}
Pandey~S, Schumacher~J, Sreenivasan~KR. A perspective on machine learning in
  turbulent flows. Journal of Turbulence. 2020;\hspace{0pt}0(0):1--18.
  \urlprefix\url{https://doi.org/10.1080/14685248.2020.1757685}.

\bibitem{Thompson2016}
Thompson~RL, Sampaio~LEB, de~Braganca~Alves~F, et~al. A methodology to evaluate
  statistical errors in dns data of plane channel flows. Comput Fluids.
  2016;\hspace{0pt}130:1--7.

\bibitem{Almeida1993}
Almeida~G, Durao~D, Heitor~M. Wake flows behind two-dimensional model hills.
  Experimental Thermal and Fluid Science. 1993;\hspace{0pt}7(1):87--101.

\bibitem{Mellen2000}
Mellen~C, Froehlich~J, Rodi~W. Large eddy simulation of the flow over periodic
  hills. In: In: Proceedings. 16th IMACS World Congress, Lausanne, Switzerland;
  2000.

\bibitem{Breuer2009}
Breuer~M, Peller~N, Rapp~C. Flow over periodic hills: Numerical and
  experimental study in a wide range of reynolds numbers. Comput Fluids.
  2009;\hspace{0pt}38:433--457.

\bibitem{Xiao2019}
Xiao~H, Wu~JL, Laizet~S, et~al. Flows over periodic hills of parameterized
  geometries: A dataset for data-driven turbulence modeling from direct
  simulations ; 2019.

\bibitem{Temmerman2003}
Temmerman~L, Leschziner~MA, Mellen~CP, et~al. Investigation of wall-function
  approximations and subgrid-scale models in large eddy simulation of separated
  flow in a channel with streamwise periodic constrictions. Int J Heat Fluid
  Flow. 2003;\hspace{0pt}24(2):157--180.

\bibitem{Xia2013}
Xia~Z, Shi~Y, Hong~R, et~al. Constrained large-eddy simulation of separated
  flow in a channel with streamwise-periodic constrictions. J~Turbul.
  2013;\hspace{0pt}14(1):1--21.

\bibitem{Wu2019a}
Wu~JL, Sun~R, Laizet~S, et~al. Representation of stress tensor perturbations
  with application in machine-learning-assisted turbulence modeling. Comput
  Methods Appl Mech Engrg. 2019;\hspace{0pt}346:707--726.

\bibitem{Caretto1973}
Caretto~LS, Gosman~AD, Patankar~SV, et~al. Two calculation procedures for
  steady, three-dimensional flows with recirculation. In: Cabannes~H, Temam~R,
  editors. Proceedings of the Third International Conference on Numerical
  Methods in Fluid Mechanics; Berlin, Heidelberg. Springer Berlin Heidelberg;
  1973. p. 60--68.

\bibitem{OpenFOAM}
Weller~HG, Tabor~G, Jasak~H, et~al. A tensorial approach to computational
  continuum mechanics using object-oriented techniques. Comput Physics.
  1998;\hspace{0pt}12(6):620--631.

\bibitem{kOmegaSST}
Menter~F, Kuntz~M, Langtry~R. Ten years of industrial experience with the sst
  turbulence model. Heat and Mass Trans. 2003 01;\hspace{0pt}4:625--632.

\bibitem{AdaGrad}
Duchi~J, Hazan~E, Singer~Y. Adaptive subgradient methods for online learning
  and stochastic optimization. J Machine Learning Res. 2011
  07;\hspace{0pt}12:2121--2159.

\bibitem{Adam}
Kingma~DP, Ba~J. Adam: A method for stochastic optimization ; 2014.

\end{thebibliography}

\end{document}